\documentclass{svproc}
\usepackage{url}

\usepackage{amsmath}
\usepackage{algorithm}
\usepackage[noend]{algpseudocode}
\usepackage{multicol}
\usepackage{graphicx}
\usepackage{subfig}
\usepackage{setspace}
\usepackage{cite}

\algdef{SxnE}[ENUM]{Enum}{EndEnum}[1]{\algorithmicenums}
\algnewcommand{\algorithmicenums}{\textbf{enum }}

\algdef{SxnE}[STRUCT]{Struct}{EndStruct}{\algorithmicstructs}
\algnewcommand{\algorithmicstructs}{\textbf{struct }}

\begin{document}
\mainmatter              
\title{Lock-Free Transactional Adjacency List\protect\footnote{This research was supported by the National Science Foundation under NSF OAC 1440530, NSF CCF 1717515, and NSF OAC 1740095.}}
\titlerunning{Transactional Adjacency List}  
%
\author{Zachary Painter, Christina Peterson, Damian Dechev}
\authorrunning{} 
%
\tocauthor{}
\institute{University Of Central Florida, Orlando FL 32816, USA}

\maketitle              

\begin{abstract}
Adjacency lists are frequently used in graphing or map based applications. Although efficient concurrent linked-list algorithms are well known, it can be difficult to adapt these approaches to build a high-performance adjacency list. Furthermore, it can often be desirable to execute operations in these data structures transactionally, or perform a sequence of operations in one atomic step. In this paper, we present a lock-free transactional adjacency list based on a multi-dimensional list (MDList). We are able to combine known linked list strategies with the capability of the MDList in order to efficiently organize graph vertexes and their edges. We design our underlying data structure to be node-based and linearizable, then use the Lock-Free Transactional Transformation (LFTT) methodology to efficiently enable transactional execution. In our performance evaluation, our lock-free transactional adjacency list achieves an average of 50\% speedup over a transactional boosting implementation. 
\end{abstract}
\section{Introduction}
Lock-free data structures aim to fully utilize the computing resources of multi-core processors without the drawbacks of lock-based counterparts such as deadlock or priority inversion. However, lock-free data structures are difficult to design due to the consideration of all possible thread interleavings when reasoning about safety or liveness properties. Even more so are lock-free transactional  data structures because in addition to the safety and liveness properties of traditional lock-free data structures, isolation must be preserved such that a series of operations appear to occur in one atomic step.

An adjacency list data structure maps graph nodes, or ``vertexes," to other nodes by their connections, or ``edges." Generally, if a vertex $i$ is adjacent to another vertex $j$, then vertex $j$ is contained in the sublist of vertex $i$. In order to implement such a data structure concurrently, one would need to overcome the challenges of traversing in multiple dimensions, organizing vertex and edge nodes, and properly disposing of all children of a vertex before deleting the vertex.

Previous work on lock-free linked list data structures are designed for sets and queues. Since elements of these abstract data types do not account for relationships between elements, they are unsuitable to be directly used for an adjacency list data structure. An adjacency list data structure needs to support operations that can insert and remove vertexes and edges, as well as check whether a vertex or edge is contained in the list. Additional synchronization is required to ensure that an operation that deletes a vertex $i$ does not modify or remove nodes that are currently part of $i$'s sublist of adjacent nodes. Further synchronization is required to ensure that two operations are able to simultaneously modify the sublist of a vertex despite those operations appearing to take place at the same vertex.

A lock-free adjacency list provides atomicity at the granularity of an individual operation. However, in some cases one may want to perform a sequence of operations such that the entire sequence appears to take place in one atomic step. One such case is during the deletion of a vertex, in which case it must first be guaranteed that all edges from that vertex have already been deleted. In such a case, a sequence of operations such as the following would be useful.

\vspace{-1em}
\begin{algorithm}
	\begin{algorithmic}[1]
		\If {\textsc{isEmpty}(\textit{vertex.List})}
		\State \textsc{Delete}(\textit{vertex});
		\EndIf
	\end{algorithmic}
\end{algorithm}

\vspace{-2em}
This code should be able to verify that a given node's sublist is empty before deleting that node. Unfortunately, this operation fails to complete its goal. Since the composition of the methods is not atomic, another thread $a$ could insert an edge between the time that thread $b$ reads that the list of edge nodes is empty, and thread $b$ deleting the vertex, thus invalidating the operation.

In order to perform a series of operations such as those previously mentioned, all involved operations need to appear to take place in a single atomic step. Additionally, if any operation fails, it must appear as though none of the operations took place. Some implementations, such as Transactional Boosting \cite{Maurice}, use fine-grained locking in order to create a transactional data structure from an underlying concurrent data structure. This, however, reduces the performance of the data structure, and negates any lock-free progress guarantee  the underlying data structure might have had. Software Transactional Memory (STM) can also be used to create transactional data structures from existing ones. Unfortunately, this approach also creates significant performance loss. In an STM data structure, transactions maintain a list of read and write locations. If a transaction's read and write set overlaps with another transaction's write set, those transactions conflict. In the case of a conflict, one of the transactions must abort. This results in a significant amount of unnecessary aborts, as conflicts detected in this way do not necessarily correspond to high-level semantic conflicts. These excessive aborts can severely limit the degree of concurrency when executing transactions on a data structure.

In this paper, we present a high performance lock-free transactional adjacency list. The primary goal of the data structure presented in this work is to (1) implement a lock-free adjacency list base data structure, and (2) enable transactional execution of operations in this data structure.

\begin{figure}
\centering
\includegraphics[trim={0cm 5cm 0cm 0cm},clip=true,scale=0.45]{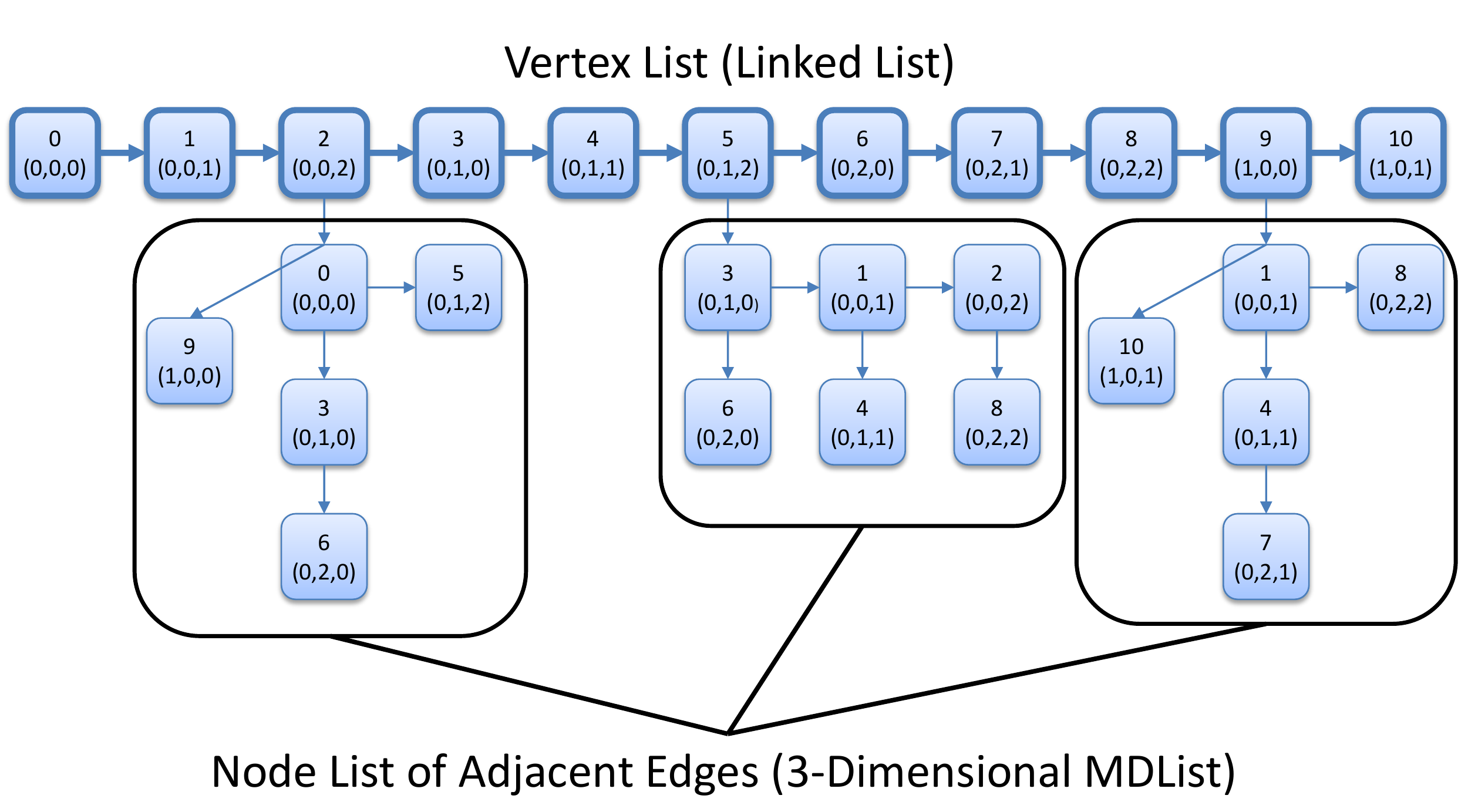}
\caption{Adjacency List Structure} \label{Fig:AdjacencyList}
\end{figure}

In order to achieve the first goal, we implement lock-free adjacency list using a lock-free linked list of vertexes, where each vertex contains a pointer to a Multi-Dimensional List (MDList)~\cite{Zhang2} to allow fast lookup of edges. 
We depict the adjacency list structure in Figure~\ref{Fig:AdjacencyList}.
An MDList guarantees a worst-cast search time complexity of $O(\log{}N)$, an improvement over a worst-cast search time complexity of $O(N)$ provided by design alternatives such as a linked list or skiplist.
A skiplist provides an average search time complexity of $O(\log{}N)$, but has a worst-cast search time complexity of $O(N)$ if shortcuts to the node of interest do not exist.
We place all vertexes in the primary linked list, and all adjacent edges to that vertex as a node in its associated MDList. This allows us to take maximum advantage of the multi-dimensional property of the MDList, while also easily organizing the relative locations of each vertex and their corresponding edges. Background details on the MDList are provided in Section~\ref{background}. 

We refer to elements in the primary linked list as vertexes, and elements in the sublist of a vertex as nodes. A node $a$ contained in vertex $b$'s associated MDList indicates that vertex $a$ is adjacent to vertex $b$. When inserting or deleting a vertex, we traverse along the main list of vertexes, checking each key, until we find the location to insert or delete our vertex. While allocating the vertex we also allocate a new MDList for that vertex to point to.

\vspace{-2em}
\vspace{-2em}

In order to achieve the second goal, we adopt Lock-Free Transaction Transformation (LFTT) \cite{Zhang1} by storing descriptor objects within each node in both the main list and each MDList. LFTT uses high-level semantic conflict detection to avoid low-level read/write conflicts, and a logical rollback to avoid the performance penalties of a physical rollback. Background details on LFTT are provided in Section~\ref{background}.

The contribution made by this paper is as follows:

\begin{itemize}
	\item To the best of our knowledge, this paper presents the only lock-free transactional adjacency list.
\end{itemize}

\begin{itemize}
	\item This data structure experiences an average speedup greater than 50\% when compared to similar approaches based on transactional boosting and STM. 
\end{itemize}

\section{Background}
\label{background}
An MDList partitions a linked list into shorter lists organized in multi-dimensional space to improve search time. A node in a $D$-dimensional MDList comprises a key-value pair, a coordinate vector of integers $k[D]$, and an array of child pointers where the $d$th pointer links to a child node of dimension $d$. A list of arbitrary dimension $D$ is formally defined as follows.

\begin{definition}
A $D$-dimensional list is a rooted tree in which each node is implicitly assigned a dimension of $d \in [0,D)$. The root node's dimension is 0. A node of dimension $d$ has no more than $D - d$ children, and each child is assigned a unique dimension of $d' \in [d,D)$ \cite{Zhang2}. 
\end{definition}

Given a key range of $[0,N)$ in a $D$-dimensional space, the maximum number of keys in each dimension is $b = \lceil  ~ \sqrt[D]{N}~ \rceil$. The mapping of an integer key to its $D$-dimension vector coordinates is performed by converting the key to a $b$-based number and using each digit as an entry in the vector coordinates. Each node is associated with a coordinate vector $k$, where a dimension $d$ node shares a coordinate prefix of length $d$ with its parent. The following definition provides the criteria for which nodes are ordered in their $D$-dimensional list.

\begin{definition}
Given a non-root node of dimension $d$ with coordinate \\$k = (k_0, ...,k_{D-1})$ and its parent with coordinate $k'=(k'_0, ..., k'_{D-1})$ in an ordered D-dimensional list: $k_i = k'_i, \forall i \in [0,d) \wedge k_d > k'_d$ \cite{Zhang2}.
\end{definition}

The search for a node is performed by starting at the 0-dimension and traversing all nodes at this dimension until either a node with the same 0th coordinate as the key of interest is reached, or the current node being traversed has a greater 0th coordinate than the key of interest. If a node with a 0th coordinate identical to the key of interest exists, then the search advances to the next dimension $d$. The search will continue advancing dimensions given that a node with the same $d$th coordinate as the key of interest is found. The search terminates when either a node with the same coordinates as the key of interest is found, or no node exists with the same $d$th coordinate as the key of interest. 

The worst-case time complexity of a search in an MDList is $O(D \cdot b)$, where $b$ is the maximum number of nodes in a dimension. Replacing $b$ in the worst-cast time complexity, we have $O(D \cdot b) = O(D \cdot  \sqrt[D]{N})$. If we choose $D \propto \log{}N$, then $ O(D \cdot  \sqrt[D]{N}) = O(\log{}N \cdot  \sqrt[\log{}N]{N}) =  O(\log{}N \cdot 2) = O(\log{}N) $.
 
Insertion into an MDList is performed by splicing and child adoption. Splicing consists of updating the new node's child pointer to point to the predecessor's child, and updating the predecessor's child pointer to point to the new node. Child adoption is necessary when the dimension of an old child has changed due to the insertion of a new node, where the old child will be adopted as a higher dimension child of the new node. Deletion of a node in an MDList is performed by updating the predecessor's child point to point to the child of the node to be deleted. In the case of a deletion, the child of the node to be deleted is adopted as a lower dimension child of the predecessor.

Lock Free Transactional Transformation (LFTT) is a methodology for creating transactional data structures from lock-free node-based data structures. LFTT handles conflicts between operations by utilizing descriptor objects referenced by each node. These transaction descriptors contain all information necessary for an arbitrary thread to perform any given operation or sequence of operations belonging to a transaction. For a thread to perform an operation at a node as part of a transaction, it is must first create a reference to its transaction descriptor in the node. If there already exists a transaction descriptor at that node, a conflict between two transactions accessing the same node has been detected. LFTT resolves these conflicts by having the thread that finds an existing transaction descriptor at a node help complete the conflicting transaction by executing all remaining operations that are part of that transaction, thus eventually causing the conflicting transaction to either succeed or fail. Once the transaction referenced by the transaction descriptor at a node is complete, a thread may place a reference to its own transaction descriptor in the node and proceed with its operation. 

LFTT additionally handles the recovery of failed transactions through its transaction descriptors. A transaction descriptor may be marked as $committed$, indicating that all operations that are part of the transaction have been successfully completed. Alternatively, a transaction descriptor may be marked $aborted$, indicating that none of the operations in the transaction should occur. LFTT is able to avoid the need to physically undo already completed operations that are part of an aborted transaction by interpreting the logical status of a node based on its transaction descriptors status. A nodes status in the list is interpreted inversely if it is part of an aborted transaction. This results in the \textit{appearance} that all completed operations that are part of an aborted transaction have been undone.  

\vspace{-1em}
\section{Lock-Free Transactional Adjacency List}

The primary challenge in creating a lock-free transactional adjacency list is its multi-dimensional structure, which poses a major challenge to performing transactional synchronization for non-commutative operations. \textsc{InsertEdge} and \textsc{DeleteEdge} create a relation between two vertexes by adding or removing a node from the sublist of an existing vertex. Any \textsc{InsertEdge} or \textsc{DeleteEdge} operation occurring at vertex $j$ would have their outcome affected by a transaction that modifies vertex $j$. As a result, two edge operations occurring at the same vertex are able to commute, while an edge operation and an operation that modifies the vertex itself are not. The \textsc{DeleteVertex} method requires special consideration. The case in which a transaction deletes a vertex at which one or multiple threads are performing an edge operation must be prevented. A \textsc{DeleteVertex} operation on a vertex should help complete all pending edge operations currently accessing the MDList contained at that vertex. Simultaneously, any subsequent operations attempting to access that MDList will first help complete the pending \textsc{DeleteVertex}.

The constants provided by LFTT are detailed in Algorithm \ref{Def}. We introduce a \textit{currentOp} field to each descriptor to track the current progress of each transaction. The \textsc{IsNodePresent, IsKeyPresent, ExecuteOps, MarkDelete, LocatePred}, and pointer marking operations are provided in Lock-Free Transactional Transformation \cite{Zhang1}.

\vspace{-1.5em}
\begin{algorithm}
	\caption{LFTT Definitions}\label{Def}
	\scriptsize
	\begin{multicols}{2}
		\begin{algorithmic}[1]
			
			\Enum \textbf{TxStatus}
			\State Active
			\State Committed
			\State Aborted\
			\EndEnum
			
			\Enum \textbf{OpType}
			\State InsertVertex
			\State DeleteVertex
			\State InsertEdge
			\State DeleteEdge
			\State Find
			\EndEnum
			
			\Struct \textbf{Operation}
			\State \textbf{OpType} type
			\State \textbf{int} key 
			\EndStruct
			
			\Struct \textbf{Desc}
			\State \textbf{int} size
			\State \textbf{TxStatus} status
			\State \textbf{int} currentOp
			\State \textbf{Operation} ops[] 
			\EndStruct
			
			\Struct \textbf{NodeDesc}
			\State \textbf{Desc*} desc
			\State \textbf{int} opid
			\EndStruct
			
			\Struct \textbf{Node}
			\State \textbf{NodeDesc*} info
			\State \textbf{int} key
			\State \textbf{MDList*} list
			\State ...
			\EndStruct
			
		\end{algorithmic}
	\end{multicols}
\end{algorithm}

\vspace{-3em}
\begin{algorithm}
	\caption{Update Info Pointer}\label{Update}
	\scriptsize
	\begin{algorithmic}[1]
		
		\Function{UpdateInfo}{\textbf{Node*} \textit{n}, \textbf{NodeDesc* }\textit{info}, \textbf{bool} \textit{wantkey}}
		
		\State \textbf{NodeInfo }*\textit{oldinfo $\leftarrow$ n.info}
		
		\If{\textsc{IsMarked}(\textit{oldinfo})} 
		\State \textsc{Do\_Delete($n$)} 
		\State \Return retry;
		\EndIf
		
		\If{\textit{oldinfo.desc $\neq$ info.desc}}
		\If{\textit{oldinfo.desc.ops[oldinfo.opid] == DeleteVertex \& oldinfo.desc.currentOp == oldinfo.opid}}
		\State \textsc{ExecuteOps}(\textit{oldinfo.desc, oldinfo.opid})
		\Else
		\State \textsc{ExecuteOps(\textit{oldinfo.desc, oldinfo.opid+1})}
		\EndIf
		\ElsIf {\textit{oldinfo.opid} $>=$ \textit{info.opid}}
		\State \Return success
		\EndIf
		
		\State \textit{haskey} $\leftarrow$ \textsc{IsKeyPresent}(\textit{oldinfo})
		
		\If{(!\textit{haskey} \& \textit{wantkey}) $||$ (\textit{haskey} \& !\textit{wantkey})}
		\State \Return fail
		\EndIf
		
		\If{\textit{info.desc.status $\neq$ Active}}
		\State \Return fail
		\EndIf
		
		\If{CAS(\textit{\&n.info, oldinfo, info})}
		\State \Return success
		\Else
		\State \Return retry
		\EndIf

		\EndFunction
		
	\end{algorithmic}
\end{algorithm}

Algorithm \ref{Update} contains the \textsc{UpdateInfo} operation provided by Lock-Free Transactional Transformation \cite{Zhang1}, which has been modified in the following way to allow for a special case regarding \textsc{DeleteVertex}. At line 2.6 we check the info pointer at node \textit{n}. If a different operation is currently taking place at node \textit{n}, that operation must be completed before the desired operation can begin. At line 2.7 we check if the operation that occurred at node \textit{n} was a \textsc{DeleteVertex} operation. If so, we check whether the \textsc{DeleteVertex} operation is pending. The \textit{currentOp} variable stores what step the transaction is currently on. If this value is equal to the value of the operation that occurred at node \textit{n}, then the \textsc{DeleteVertex} operation is not complete and the current thread should use the descriptor object to attempt to delete the vertex. For all other operations, the presence of an info pointer at node \textit{n} indicates that the operation described by \textit{n.info} is already complete. Thus, \textsc{ExecuteOps} is called on the next operation in the transaction. 

\begin{algorithm}
	\caption{Transformed Delete Vertex}\label{DelVert}
	\scriptsize
	\begin{algorithmic}[1]
		
		\Function{DeleteVertex}{\textbf{int} \textit{vertex}, \textbf{NodeDesc* }\textit{nDesc}}
		
		\State \textbf{Node }*\textit{curr $\leftarrow$ head}
		\State \textbf{Node }*\textit{pred $\leftarrow$ NULL}
		
		\While{true}
		
		\State \textsc{LocatePred}(\textit{pred, curr, vertex})
		
		\If{\textsc{IsNodePresent}(\textit{curr, vertex})}
		\State \textit{ret} $\leftarrow$ \textsc{UpdateInfo}(\textit{curr, nDesc, true})
		
		\If{\textit{ret}}
		\State \textit{MDList *list $\leftarrow$ curr.list}
		\State \textit{ret$\leftarrow$list}.\textsc{FinishDelete}(\textit{list.head, 0, nDesc})
		\EndIf
		\Else
		\State \textit{ret $\leftarrow$ false}
		\EndIf
		
		\If{\textit{ret}}
		\State \Return \textit{true}
		\Else
		\State \Return \textit{false}
		\EndIf
		
		\EndWhile
		
		\EndFunction
		
	\end{algorithmic}
\end{algorithm}

\vspace{-3em}
\begin{algorithm}
	\caption{Finish Pending \textsc{DeleteVertex} Operation}\label{FinishDel}
	\scriptsize
	\begin{algorithmic}[1]
		
		\Function{FinishDelete}{\textbf{MDList::Node* } \textit{n}, \textbf{int} \textit{dc}, \textbf{NodeDesc* } \textit{nDesc}}
		
		\While {true}
		
		\If{\textsc{UpdateInfo}(\textit{n, nDesc, true})}
		\State \textbf{Break}
		\EndIf
		
		\EndWhile
		
		\For{i $\in$ [dc,DIMENSION)}
		\State \textbf{MDList::Node } *\textit{child} $\leftarrow$ \textit{n.child[i]}
		\State \textsc{CAS}(\&\textit{n.child[i], child}, \textsc{Set\_Mark}(\textit{child}))
		
		\If{$child != NULL$}
		\State ret $\gets$ \textsc{FinishDelete}($child, i, nDesc$)
		\If {ret == $false$}
		\State \textbf{return} false
		\EndIf
		\Else
		\State \textbf{return} false
		\EndIf
		\EndFor
		
		\EndFunction
		
	\end{algorithmic}
\end{algorithm}

\vspace{-2em}
\subsection{Adjacency List Operations}
This adjacency list supports 5 operations: \textsc{InsertVertex, DeleteVertex, InsertEdge, DeleteEdge,} and \textsc{Find}. The \textsc{InsertVertex} operation adds a vertex to a primary linked list of vertexes. The \textsc{InsertEdge} operation adds a node to a specific vertex's sublist, thus establishing that node as adjacent to the specified vertex. The \textsc{DeleteVertex} and \textsc{DeleteEdge} operations are the inverses of their counterparts. The \textsc{Find} operation searches for a node within the sublist of vertex $j$, returning whether or not that node shares an edge with vertex $j$.

Algorithm \ref{DelVert} details the \textsc{DeleteVertex} operation. \textsc{DeleteVertex} traverses the main list of vertexes by calling \textsc{LocatePred} on line 3.5. If the node with the target key already exists, then \textsc{LocatePred} will return when $curr$ points to the node with that key, otherwise, $curr$ will point to the logical successor of the node to be deleted. We check for the case that the node with the desired key already exists on line 3.6. We then call \textsc{UpdateInfo} to attempt to redirect the $info$ pointer. If this succeeds, we must then call \textsc{FinishDelete} on the vertex's $list$ object. \textsc{FinishDelete} traverses $list$ calling \textsc{UpdateInfo} on all the nodes it contains. Additionally, we must mark the next pointer of all nodes as they are traversed, which will interrupt competing \textsc{InsertEdge} operations that have already begun inserting their node on line 6.15, causing them to re-traverse. The goal of this operation is to logically delete all edges adjacent to the vertex to be deleted. This process allows all pending transactions occurring within the sublist to commit due to the call to \textsc{UpdateInfo} at line 4.3. Once it can be guaranteed that all nodes within the vertex's list are deleted, the operation is complete. Physical deletion is later done by using Compare-And-Swap to change $pred.next$ to point to $curr.next$, thus removing the vertex from the main list. 

The \textsc{InsertVertex} algorithm is similar to \textsc{DeleteVertex}. \textsc{InsertVertex} traverses the list using \textsc{LocatePred}, but can only succeed if its value is not already in the list (\textsc{!IsNodePresent}($curr, vertex$)). In this case, it allocates a new vertex and inserts it into the list using Compare-And-Swap to change $pred.next$ to $curr$. 

\begin{algorithm}
	\caption{Find Vertex Operation}\label{FindVert}
	\scriptsize
	\begin{algorithmic}[1]
		
		\Function{FindVertex}{\textbf{int} \textit{vertex}, \textbf{NodeDesc* }\textit{nDesc}, \textbf{int} \textit{opid}}
		
		\State \textbf{Node }*\textit{curr $\leftarrow$ head}
		\State \textbf{Node }*\textit{pred $\leftarrow$ NULL}
		
		\While{true}
		
		\State \textsc{LocatePred}(\textit{pred, curr, vertex})
		
		\If{\textsc{IsNodePresent}(\textit{curr, vertex})}
		\State \textbf{NodeDesc }*\textit{cDesc $\leftarrow$ curr.info}
		\If{\textit{cDesc != nDesc}}
		\State \textsc{ExecuteOps}(\textit{cDesc.desc,cDesc.opid+1})
		\EndIf
		
		\If{\textsc{IsKeyPresent}(\textit{cDesc})}
		
		\If{\textit{nDesc.desc.status != ACTIVE}}
		\State \Return \textit{NULL}
		\Else
		\State \Return \textit{curr}
		\EndIf
		
		\EndIf
		
		\Else
		\State \Return \textit{NULL}
		\EndIf
		
		\EndWhile
		
		\EndFunction
		
	\end{algorithmic}
\end{algorithm}

Algorithm \ref{FindVert} details the main method used to help the insertion of edge nodes. To begin, it searches the list until it finds the correct vertex node and verifies that it is logically in the list, and that no other transaction currently holds the $info$ pointer. If another thread does hold the $info$ pointer, the thread will help complete that transaction at line 5.9. Otherwise, the function returns a pointer to the node.

\begin{algorithm}
	\caption{Insert key:edge at target vertex}\label{InsEdge}
	\scriptsize
	\begin{algorithmic}[1]
		
		\Function{InsertEdge}{\textbf{int} \textit{vertex}, \textbf{int} \textit{edge}, \textbf{NodeDesc* }\textit{nDesc}, \textbf{int} \textit{opid}}
		
		\While{true}
		
		\State \textbf{Node } *\textit{currVertex} $\leftarrow$  \State \textsc{FindVertex}(\textit{vertex, nDesc, opid}))
		
		\If{\textit{currVertex == NULL}}
		\State \Return \textit{false}
		\EndIf
		
		\State \textbf{Node }*\textit{pred $\leftarrow$ NULL} 
		\State \textbf{Node }*\textit{currEdge $\leftarrow$ currVertex.list.head}
		
		\While{true}
		\State \textit{currVertex.list}.\textsc{LocatePred}(\textit{pred, currEdge})
		
		\If{\textsc{IsNodePresent}(\textit{currEdge, edge})}
		\State \textbf{return} \textit{false}
		\Else			
		\State \textbf{MDList::Node }*\textit{n} $\leftarrow$ \textbf{new }MDList::Node
		\State \textit{n.info} $\leftarrow$ \textit{nDesc}
		\State \textbf{return} currVertex.list.\textsc{Do\_Insert}($n$)
		\EndIf
		
		\EndWhile
		
		\EndWhile
		
		\EndFunction
		
	\end{algorithmic}
\end{algorithm} 

\begin{algorithm}
	\caption{Delete key:edge at target vertex}\label{DelEdge}
	\scriptsize
	\begin{algorithmic}[1]
		
		\Function{DeleteEdge}{\textbf{int} \textit{vertex}, \textbf{int} \textit{edge}, \textbf{NodeDesc* }\textit{nDesc}, \textbf{int} \textit{opid}}
		
		\While{true}
		
		\State \textbf{Node } *\textit{currVertex} $\leftarrow$ \State \textsc{FindVertex}(\textit{vertex, nDesc, opid}))
		
		\If{\textit{currVertex} == \textit{NULL}}
		\State \Return false
		\EndIf
		
		\State \textbf{Node }*\textit{pred} $\leftarrow$ \textit{NULL}
		\State \textbf{Node }*\textit{currEdge} $\leftarrow$ \textit{currVertex.list.head}
		
		\While{true}
		\State \textit{currVertex.list}.\textsc{LocatePred}(\textit{pred, currEdge})
		
		\If{\textsc{IsNodePresent}(\textit{currEdge, edge})}
		\State \textbf{return} (\textsc{UpdateInfo}(\textit{currEdge, nDesc, true}) == $success$)
		\Else
		\State \textbf{return} \textit{false}
		\EndIf
		
		\EndWhile
		
		\EndWhile
		
		\EndFunction
		
	\end{algorithmic}
\end{algorithm} 

Algorithms \ref{InsEdge} details the insertion of a node into an MDList in order to create an edge with a vertex. \textsc{InsertEdge} begins by calling \textsc{FindVertex} to get the proper vertex node for insertion. If the node exists, then we traverse the MDList pointed to by the vertex to find the proper location to insert the new edge node. Once the traversal is complete, insertion is done the same way as in \textsc{InsertVertex}. 

Algorithm \ref{DelEdge} details the deletion of a node in an MDList in order to remove an edge with a vertex. \textsc{DeleteEdge} traverses to the target vertex using the same logic as \textsc{InsertEdge}. Once it has acquired a valid vertex, it traverses the MDList looking for the target edge node to delete. If the target node is found in the MDList, deletion is done by updating the $info$ pointer of the target node.
\vspace{-1em}
\section{Correctness}

The lock-free transactional adjacency list is designed for the correctness property strict serializability. According to conclusion by Herlihy et al. \cite{Maurice}, a committed transaction is strictly serializable given that a data structure contains linearizable operations and obeys commutativity isolation.

\vspace{-1em}
\subsection{Definitions}

According to Herlihy et al. \cite{Maurice}, a $history$ is a sequence of instantaneous events. Events occur during the transition of a transactions status between pending, committed, and aborted. \\

\vspace{-1em}
\begin{definition}
A history h is strictly serializable if the committed series of transactions is equivalent to a legal history in which all transactions executed sequentially in the order they commit.
\end{definition}

\begin{definition}
Two method calls I,R and I',R' commute if: for all histories h, if h $\cdot$ I $\cdot$ R and h $\cdot$ I' $\cdot$ R' are both legal, then h $\cdot$ I $\cdot$ R $\cdot$ I' $\cdot$ R' and h $\cdot$ I' $\cdot$ R' $\cdot$ I $\cdot$ R are both legal and define the same abstract state.	
\end{definition}

Operations are said to commute if executing them in any order yields the same abstract state. The commutativity of adjacency list operations are as follows, assuming vertexes \textit{x,y} and nodes \textit{i,j}: \\

\begin{spacing}{1.1}
	\noindent
	\textsc{InsertVertex(\textit{x}) $\leftrightarrow$ InsertVertex(\textit{y})}, x $\neq$ y \\
	\textsc{DeleteVertex(\textit{x}) $\leftrightarrow$ DeleteVertex(\textit{y})}, x $\neq$ y \\
	\textsc{InsertVertex(\textit{x}) $\leftrightarrow$ DeleteVertex(\textit{y})}, x $\neq$ y \\
	\textsc{InsertEdge(\textit{x, i}) $\leftrightarrow$ InsertEdge(\textit{x, j})}, i $\neq$ j \\
	\textsc{InsertEdge(\textit{x, i}) $\leftrightarrow$ InsertEdge(\textit{y, i})}, x $\neq$ y \\
	\textsc{DeleteEdge(\textit{x, i}) $\leftrightarrow$ DeleteEdge(\textit{x, j})}, i $\neq$ j \\
	\textsc{DeleteEdge(\textit{x, i}) $\leftrightarrow$ DeleteEdge(\textit{y, i})}, x $\neq$ y \\
	\textsc{InsertEdge(\textit{x, i}) $\leftrightarrow$ DeleteEdge(\textit{x, j})}, i $\neq$ j \\
	\textsc{InsertEdge(\textit{x, i}) $\leftrightarrow$ DeleteEdge(\textit{y, i})}, x $\neq$ y \\
	\textsc{FindVertex(\textit{x}) $\leftrightarrow$ InsertVertex(\textit{x})/\textit{false} $\leftrightarrow$ DeleteVertex(\textit{x})/\textit{false}}
	\\
	\textsc{FindEdge(\textit{x, i}) $\leftrightarrow$ InsertEdge(\textit{x, i})/\textit{false} $\leftrightarrow$ DeleteEdge(\textit{x, i})/\textit{false}}
	\\
\end{spacing}

\noindent
\textbf{Rule 1. \textit{Linearizability}}
\textit{For any history h, two concurrent invocations I and I' must be equivalent to either the history h $\cdot$ I $\cdot$ R $\cdot$ I' $\cdot$ R' or the history h $\cdot$ I' $\cdot$ R' $\cdot$ I $\cdot$ R}\smallbreak

\noindent
\textbf{Rule 2. \textit{Commutativity Isolation:}} 
\textit{For any non-commutative method calls $I_1$,$R_1$ $\in$ $T_1$ and $I_2$,$R_2$ $\in$ $T_2$, either $T_1$ commits or aborts before any additional method calls in $T_2$ are invoked, or vice-versa.}\smallbreak

To meet the specifications of the correctness condition linearizability, we identify an operation's linearization points. Furthermore, we will identify an operation's decision points and state-read points. The decision point of an operation occurs the moment the outcome of the operation is decided atomically. A state-read point occurs when the deciding state of the data structure is read. 

\noindent
\textbf{Lemma 1.}
\textit{The adjacency list operations \textsc{InsertVertex}, \textsc{DeleteVertex}, \textsc{InsertEdge}, \textsc{DeleteEdge}, and \textsc{Find} are linearizable.}\\

\noindent
\textit{Proof.} In the \textsc{DeleteVertex} operation, execution can branch at multiple points. Beginning at 3.6, if the vertex to be deleted is not found, the operation returns a \textit{fail} status. The state-read point of this execution occurs during traversal, when the thread reads $pred.next$ and does not find a node with the desired key. If the vertex is successfully found, but the operation returns $fail$ at line 2.15 or 2.17, then the state-read point occurs when $oldinfo.desc.status$ and $info.desc.status$ are read, respectively. Following a successful logical status update, the decision point is when the CAS operation at line 2.18 succeeds. 

The code path for \textsc{FinishDelete}, in which all nodes in the vertex's sublist are acquired by the transaction, is identical to the code path followed by \textsc{DeleteVertex} because of the call to \textsc{UpdateInfo} at line 4.3. Thus, the state-read and decision points for \textsc{FinishDelete} are the same as the respective cases in \textsc{DeleteVertex}. The code path for the physical deletion of the vertex is linearizable because \textsc{Do\_Delete}, which is provided by the base data structure, is linearizable. 

The same reasoning applies to the \textsc{InsertVertex, InsertEdge, DeleteEdge} and \textsc{Find} operations because they share the same \textsc{UpdateInfo} procedure for updating the logical status of a node. 

\noindent
\textbf{Lemma 2.}
\textit{The adjacency list operations \textsc{InsertVertex}, \textsc{DeleteVertex}, \textsc{InsertEdge}, \textsc{DeleteEdge}, and \textsc{Find} satisfy the commutativity isolation rule.}\\

\noindent
\textit{Proof.} As previously shown, commuting operations are those that access different vertexes, or those that access different nodes within the same vertex so long as no operation is operating on that vertex. This means that commuting operations must either operate on different vertexes or operate on different nodes rooted at the same vertex without operating on the vertex itself. Let $T_1$ denote a transaction that currently accesses vertex \textit{n1}. If another transaction $T_2$ were to access \textit{n1}, it must first perform \textsc{ExecuteOps} for $T_1$ which will either commit or abort $T_1$ before it is finished executing. Alternatively, let $T_1$ denote a transaction that currently accesses node \textit{m1} stored in the sublist of vertex \textit{n1}. If a transaction $T_2$ were to try to access vertex \textit{n1} it would first perform \textsc{ExecuteOps} for $T_1$ when it traverses to node \textit{m1} during the call to \textsc{FinishDelete} at 2.10, which will either commit or abort $T_1$. 

\noindent
\textbf{Theorem 1.}
\textit{The transformed lock-free adjacency list is strictly serializable}

\noindent
\textit{Proof.} Following Lemma 1 and Lemma 2, we can claim that the lock-free adjacency list is strictly serializable due to the conclusions by Herlihy and Koskinen\cite{Maurice}.

\vspace{-1em}
\section{Experimental Evaluation}
We compare the scalability and performance of our lock-free transactional adjacency list to related approaches based on transactional boosting\cite{Maurice} and NOrec Rochester Software Transactional Memory\cite{Marathe}. We create a related approach using transactional boosting by converting the lock-free transactional adjacency list's base data structure operations transaction boosting methodology. Additionally, an undo log is maintained per-thread for rollbacks in the boosted implementation. 

We evaluate the performance of these implementations using varying compositions of adjacency list operations. The compositions of operations are selected to highlight ``vertex" operations and ``edge" operations separately, as well measuring the effects of non-commutative or expensive operations like \textsc{DeleteVertex}. Each test consists of a series of fixed-size transactions made up of \textsc{InsertVertex, DeleteVertex, InsertEdge, DeleteEdge} and \textsc{Find} operations on random keys. The tests are performed on two systems; a 64-core NUMA system containing 4 AMD opteron 6272 16 core CPUs @2.1 GHz, and a 12-core system containing an Intel(R) Xeon(R) CPU E5-2697 v2 @ 2.7 ghz.

\vspace{-3em}
\begin{figure}
	\subfloat[40\% InsertVertex, 40\% DeleteVertex, 10\% InsertEdge, 10\% DeleteEdge]{\includegraphics[width = 0.5\linewidth]{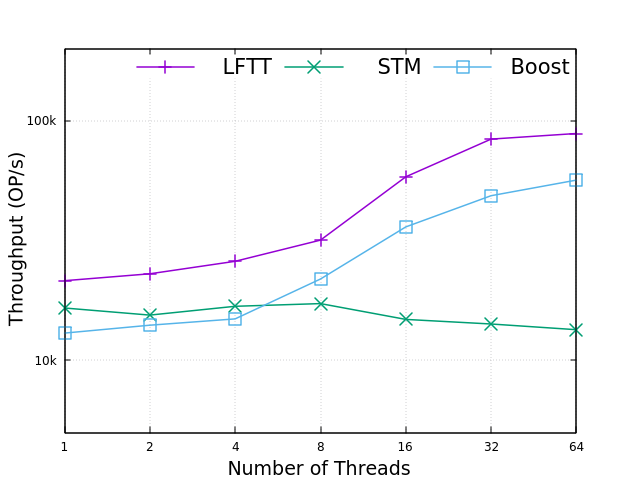}}
	\subfloat[20\% InsertVertex, 20\% DeleteVertex, 25\% InsertEdge, 25\% DeleteEdge, 10\% Find]{\includegraphics[width = 0.5\linewidth]{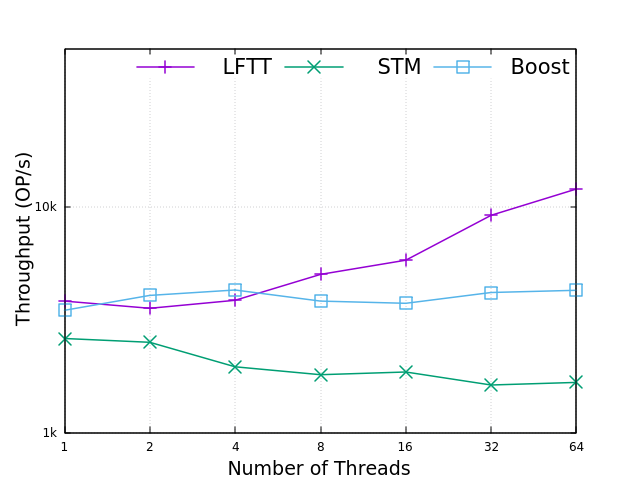}}
	\caption{NUMA System Results}
	\label{Fig:fig1}
\end{figure}

\vspace{-4em}
\begin{figure}
	\subfloat[40\% InsertVertex, 40\% DeleteVertex, 10\% InsertEdge, 10\% DeleteEdge]{\includegraphics[width = 0.5\linewidth]{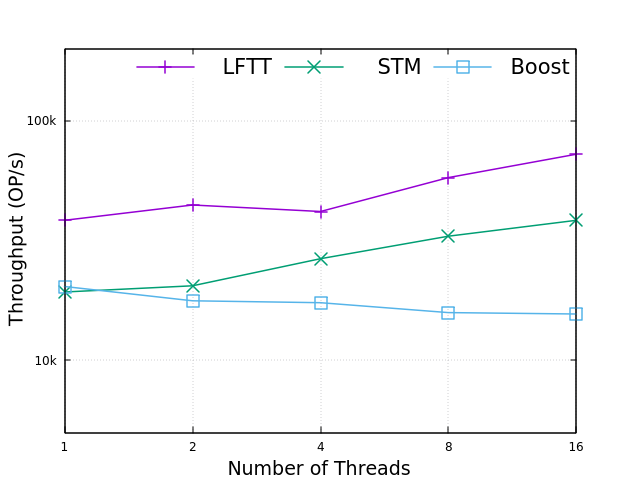}}
	\subfloat[20\% InsertVertex, 20\% DeleteVertex, 25\% InsertEdge, 25\% DeleteEdge, 10\% Find]{\includegraphics[width = 0.5\linewidth]{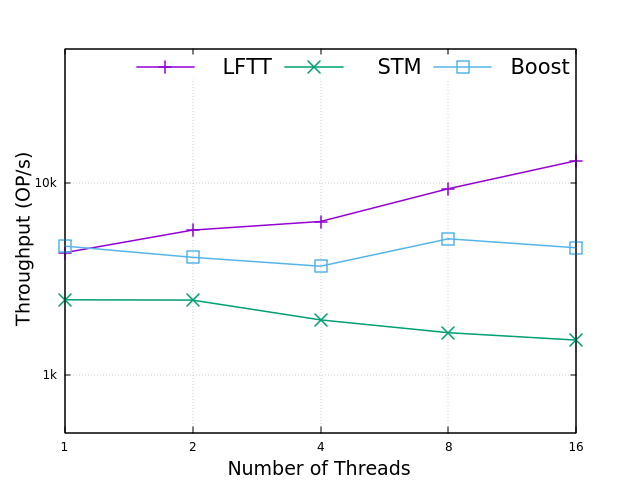}}
	\caption{Intel Xeon Results}
	\label{Fig:fig2}
\end{figure}

\vspace{-1em}
Figure \ref{Fig:fig1} shows the performance results of the 64-core NUMA system. Figure \ref{Fig:fig2} shows the results for the 12-core system. Throughput is measured in terms of operations per second. Only operations that are part of a committed transactions are counting in the calculation of throughput in order to measure the performance impact of various conflict detection and rollback schemes. The x-axis represents the number of threads running the test. In each figure, graph (a) shows a work-load dominated by operations occurring at vertexes, whereas graph (b) represents a work-load made up of relatively more operations occurring at edges. This test measures the performance impact of non-commutative operations such as \textsc{DeleteVertex} and \textsc{InsertEdge} as well as the performance impact of rollbacks on lengthy operations such as \textsc{DeleteVertex}. Each thread executed 20,000 transactions with a key range of 500. 

In Figure \ref{Fig:fig1}, the difference between the lock-free transactional adjacency list, denoted `LFTT,' the transactional boosting implementation, denoted `Boost,' and the Software Transactional Memory implementation, denoted `STM' is shown. In the boosting implementation, threads must acquire locks on nodes for each operation. In the case of \textsc{DeleteVertex}, threads may need to acquire a number of locks equal to the size of the vertex's sublist. In this case, the lock-free algorithm has the advantage of only needing to allocate a single descriptor object for the entire transaction. Additionally with regards to Boost, the cost of rolling back aborted operations is very high in operations like \textsc{DeleteVertex}. Not only must the vertex be restored after an aborted transaction, but all nodes from the vertex's sublist must be re-added using \textsc{InsertEdge}. This creates a very low performance for aborted transactions in transactional boosting. Because of LFTT's logical status update, the lock-free transactional adjacency list is able to rollback these operations in a single atomic step. Similarly, STM experiences a heavy performance loss due to its high number of spurious aborts. STM is very likely to detect a conflict between operations like \textsc{DeleteVertex}, which modify a great number of nodes, despite there being no semantic conflict between transactions. These results are highly similar to the ones gathered from the 12-core system displayed in Figure \ref{Fig:fig2}.

In general, the lock-free transactional adjacency list outperforms transactional boosting implementation by an average of 50\%, and frequently outperforms RSTM by as much as 150\%. 

\vspace{-1em}
\section{Related Work}

Transactions can be enabled in similar data structures using related approaches such as STM or Transactional Boosting. We focus our discussion on transactional Lists and Skiplists, which provide similar node-based store and search time complexities.

\vspace{-1em}
\subsection{Transactional Memory}


Transactional memory is a programming paradigm initially proposed by Herlihy et al.~\cite{herlihy1993transactional} intended to simplify concurrent programming by allowing user-specified blocks of code to be executed in hardware, exhibiting both atomicity and isolation. Software transactional memory, proposed by Shavit et al.~\cite{shavit1997software}, was developed to facilitate transactional programming without hardware transactional memory support. Herlihy et al.~\cite{herlihy2003software} present DSTM, an application programming interface for obstruction-free STM designed to support dynamic-sized data structures. Dalessandro et al.~\cite{dalessandro2010norec} present NOrec, a low-overhead STM that utilizes a single global sequence lock shared with the transactional mutex lock system, an indexed write set, and value-based conflict detection to provide features such as livelock freedom, full compatibility with existing data structure layouts, and starvation avoidance mechanisms. Dice et al.~\cite{dice2006transactional} present Transactional Locking II (TL2), an STM algorithm that uses a novel version-clock validation to guarantee that user code operates only on consistent memory states. Other STM designs include~\cite{fraser2004practical, marathe2006lowering, saha2006mcrt}. STM implementations rely on low-level conflict detection to enable transactions. These implementations generally suffer from high spurious abort counts, making them less desirable for concurrent data structures.

Initial performance experiments were performed with Hardware Transactional Memory (HTM) by Dice et al.~\cite{dice2009early}. Intel introduced Transactional Synchronization Extensions (TSX) to the x86 instruction set architecture of the Intel 4th Generation Core\texttrademark Processors~\cite{yoo2013performance}. IBM introduced HTM in the Power ISA~\cite{cain2013robust}. Both implementations offer a best-effort HTM, which means that there is no guarantee provided that a hardware transaction will commit to memory. The disadvantage of a best-effort strategy is that HTM may experience frequent aborts due to data access conflicts, hardware interrupts, limited transactional resources, or false sharing due to unrelated variables mapping to the same cache line~\cite{intelmanual}. 

Herlihy et al.~\cite{herlihy2008transactional} present transactional boosting, a methodology for transforming highly-concurrent linearizable objects into highly-concurrent transactional objects. Transactional boosting uses a high-level semantic conflict detection to allow commutative operations in separate transactions to proceed concurrently using the thread-level synchronization of the base linearizable data structure; non-commutative operations require transaction-level synchronization through the acquisition of an abstract lock. If a transaction aborts, it recovers the correct abstract state by invoking the inverse operations recorded in the undo log. 

\vspace{-1em}
\subsection{Linked Lists}
Valois~\cite{valois1995lock},  Harris~\cite{harris2001pragmatic},  Michael~\cite{michael2002high}, and Fomitchev et al.~\cite{fomitchev2004lock} present individual algorithms for a lock-free linearizable linked list based on the Compare-And-Swap operation. Valois' algorithm addresses the problem of 1) a concurrent deletion and insertion on an adjacent cell, and 2) a concurrent deletion and deletion on an adjacent cell, by requiring that every normal node in the list have an auxiliary node with only a next field as both its predecessor and successor. The auxiliary nodes prevent the undesirable circumstance of performing an insertion or deletion on a node adjacent to a node to be deleted. Harris' algorithm uses the bit-stealing technique to logically mark a node for deletion. A lazy approach is taken for the physical deletion in which a delete operation attempts to physically delete a node once using Compare-And-Swap. If Compare-And-Swap fails, then the physical deletion is left for other threads to perform if they traverse the logically deleted node. Michael's algorithm is compatible with efficient lock-free memory management methods, including IBM freelists~\cite{treiber1986systems} and the safe memory reclamation method~\cite{michael2002safe}. Fomitchev et al.'s algorithm uses backlinks that are set when a node is deleted to allow a node to backtrack to a predecessor that is not undergoing a deletion. An MDList provides a worst-case search time complexity of $O(\log{}N)$ an improvement over the $O(N)$ worst-cast search time complexity provided by a linked list.

Transactional linked list implementations based on transactional boosting use coarse-grained locking to ensure that non-commutative method calls are never allowed to execute simultaneously. The underlying linked list algorithm's linearizability is preserved during this process to handle thread level synchronization. Rollbacks are performed by calling a method's inverse operation, which causes a performance loss for aborted transactions. Zhang and Dechev\cite{Zhang2} present a lock-free transactional linked list alongside LFTT which takes advantage of a node based conflict detection scheme to preserve the underlying algorithm's lock-freedom. This approach additionally reduces the performance hit of rollbacks by introducing a logical status update scheme capable of aborting a transaction in a single atomic step. LFTT provides transformation templates for the set abstract data type, which does not account for operations in which elements are related to each other.

\vspace{-1em}
\subsection{Skiplists and Queues}
Sundell et al.~\cite{sundell2003fast} present a lock-free priority queue based on a lock-free skiplist adapted from Lotan et al.~\cite{shavit2000skiplist}. Fomitchev et al.~\cite{fomitchev2004lock} use their lock-free linked list design~\cite{fomitchev2004lock} to implement a lock-free skiplist. Each node is augmented with a pointer to the next lower level and a pointer to the base level. Herlihy et al.~\cite{herlihy2007lock} present a lock-free skiplist based on an algorithm developed by Faser~\cite{fraser2004practical}. Skiplists eliminate global rebalancing and provide a logarithmic sequential search time on average, but the worst-case search time is linear with respect to the input size. An MDList improves upon the skiplist by providing a worst-case logarithmic sequential search time. 

Spiegelman et al.\cite{Spiegelman} presented a transactional skiplist that uses STM-like techniques combined with node locking in an attempt to reduce overhead and false aborts. Spiegelman et al. additionally present a transactional queue using a pessimistic lock-based approach. In this queue, the execution of \textsc{Enqueue} operations are deferred to the final phase of the transaction, the commit phase, in order to avoid keeping track of the current head of the queue. Meanwhile, \textsc{Dequeue} operations acquire a lock on the queue until their transaction is complete. Zhang and Dechev\cite{Zhang2} preserved lock-freedom in their algorithm by transforming a skiplist using LFTT which, again, offers a performance improvement on transaction rollbacks.

\vspace{-1em}
\section{Conclusion}
In this paper we introduced an efficient lock-free adjacency list algorithm based on MDList, then enabled transactions using the LFTT methodology. We allowed for multiple threads to concurrently modify nodes rooted at the same vertex thus increasing the amount of operations that commute. When compared to similar implementations based on related approaches, our algorithm experiences performance gains across several compositions of methods. 
%
%
\bibliography{ref}{}
\bibliographystyle{plain}
\end{document}